\documentclass[aps,prl,preprint,groupedaddress,showpacs,floatfix]{revtex4}

\usepackage{amsfonts}
\usepackage{graphicx}

\def\endprf{\hfill  {\vrule height6pt width6pt depth0pt}\medskip}

\begin{document}


\title{Revisiting the Higgs sector of the Standard Model}


\author{Marco Frasca}
\email[]{marcofrasca@mclink.it}
\affiliation{Via Erasmo Gattamelata, 3 \\ 00176 Roma (Italy)}


\date{\today}

\begin{abstract}
We analyze the Higgs sector of the Standard Model in the light of recently found exact solutions of its dynamical equation. We work in a conformal regime with zero mass term. It is seen that, provided self-interaction coupling is finite, the field acquires mass and behaves exactly as a vanilla Higgs field at the tree level. Changes are expected in the decay rates to vector bosons. We evaluate this with a quantitative analysis showing how difficult could be to observe higher excited states of the Higgs field presently. This field works like an impostor with respect to the Standard Model formulation.
\end{abstract}

\pacs{14.80.bn, 14.80.Ec}

\maketitle


The recent discovery of a scalar particle with all the properties expected for the Higgs boson, announced on 4th July last year at CERN~\cite{Aad:2012tfa,Chatrchyan:2012ufa}, settles a missing part of the Standard Model and, mostly important, gives us a definite understanding on the way fundamental particles take their masses. So, it is essential to have completely understood that what has been seen corresponds exactly to the particle that was originally postulated~\cite{Englert:1964et,Higgs:1964ia,Higgs:1964pj,Guralnik:1964eu,Higgs:1966ev,Kibble:1967sv} and introduced into the Standard Model~\cite{prl_19_1264,sm_salam}. Currently, all the data tends to agree with the characteristics of a particle of this field. The observed excitation has spin 0, positive parity and is neutral. All the decay rates seem to agree with the expectations of the Standard Model within the errors. Also, it is seen as a single particle and does not appear with other companions so far. All this concurs to a complete identification with the particle of the field postulated in the sixties. But there is a missing point yet that can make the identification quite incomplete. Presently, it is impossible to have a clear idea of the form of the self-coupling of this field and if there is effectively a mass term in its Lagrangian. No conclusion can be drawn yet about this and we have to wait until LHC upgrade will be complete and the machine turned on again.

The reason to fear an impostor is that the expected decay rates, notwithstanding the couplings can be the same, can change with respect to the Standard Model as we are going to show but being anyhow in the same ballpark of the measured ones. So, at this stage, where the we are in difficulty to achieve a measurement of the potential of the Higgs field, it is mandatory a precise measurement of such rates. Currently, we can only state that the Standard Model can comfortably sit into the data but other hypotheses are not excluded as statistical fluctuations are playing a relevant role yet. This situation will surely improve in the next few months.

The model we present here is really interesting as a working example of a Higgs-like field completely fitting the bill at the tree level. Differently from other impostor Higgs~\cite{Ellis:2013ywa}, it is more difficult to unmask as at tree level perfectly agrees with a Standard Model Higgs field. Besides, it is sound as it is based on exact classical solutions of the field equations that display massive behavior even if the equations we started from are massless or have an unphysical mass term as in a Higgs field~\cite{Frasca:2009bc}. For the sake of completeness we report such solutions here. Let us consider the general equation of motion
\begin{equation}
   -\partial^2\phi+\mu_0^2\phi+\lambda\phi^3=0.
\end{equation}
For $\mu_0^2>0$ one has
\begin{equation}
\phi(x) = \pm\sqrt{\frac{2\mu^4}{\mu_0^2 + \sqrt{\mu_0^4 + 2\lambda\mu^4}}}{\rm sn}\left(p\cdot x+\theta,\sqrt{\frac{-\mu_0^2 + \sqrt{\mu_0^4 + 2\lambda\mu^4}}{-\mu_0^2 - 
   \sqrt{\mu_0^4 + 2\lambda\mu^4}}}\right)
\end{equation}
provided that
\begin{equation}
   p^2=\mu_0^2+\frac{\lambda\mu^4}{\mu_0^2+\sqrt{\mu_0^4+2\lambda\mu^4}}
\end{equation}
being $\theta$ and $\mu$ two integration constants ($\mu$ has the dimension of energy) and ${\rm sn}$ a Jacobi elliptical function that is periodic representing a nonlinear wave-like solution. We recognize here a renormalized mass from the self-interaction. When $\mu_0^2<0$ this equation has the following exact wave-like solution
\begin{equation}
   \phi(x) =\pm v\cdot {\rm dn}(p\cdot x+\theta,i)
\end{equation}
provided that
\begin{equation}
   p^2=\frac{\lambda v^2}{2}
\end{equation}
being $v=\sqrt{\frac{2\mu_0^2}{3\lambda}}$. We see that the dispersion relation has the mass term with the right sign and we are describing oscillations around one of the selected solutions $\phi=\pm \sqrt{\frac{3}{2}}v$ as the Jacobi function ${\rm dn}$ is never zero. This is the classical description of the Brout-Englert-Higgs-Guralnik-Hagen-Kibble mechanism. For our aims, we consider a conformal Higgs sector with $\mu_0^2=0$ to be consistent with the original philosophy of the Standard Model. Then, the exact solution is 
\begin{equation}
\label{eq:sol}
\phi=\mu(2/\lambda)^\frac{1}{4}{\rm sn}(p\cdot x+\theta,i)
\end{equation}
where $\mu$ and $\theta$ are integration constants and ${\rm sn}$ is a Jacobi elliptic function. This holds provided $p^2=\mu^2\sqrt{\mu/2}$ that is the field behaves like a free massive particle notwithstanding we started from a massless field. All is needed is that $\lambda$ is finite as this kind of equation displays always massive solutions with a proper dispersion relation provided $\lambda$ is non-zero. Nonlinearities conspire to generate the mass for the field. Now, one can build a quantum field theory around these solutions and the result is that the theory admits a trivial infrared fixed point. We get an infinite set of such quantum theories all trivial~\cite{Frasca:2009bc}. We consider the generating functional
\begin{equation}
    Z[j] = N\int[d\phi] e^{i\int d^Dx
    \left[\frac{1}{2}(\partial\phi)^2-\frac{\lambda}{4}\phi^4+j\phi\right]}
\end{equation}
being $N$ a normalization constant, and we take the substitution $\phi=\phi_c+\delta\phi+O(\delta\phi^2)$ being $\phi_c$ the classical solution given in eq.(\ref{eq:sol}). We will recover the results given above and the first higher correction as it should be. After the substitution has done one has
\begin{equation}
\label{eq:zf}
    Z[j]=e^{i[S_c+\int d^Dx j\phi_c]} \int[d\delta\phi] e^{i\int d^Dx
    \left[\frac{1}{2}(\partial\delta\phi)^2
    -\frac{3}{2}\phi_c^2(\delta\phi)^2+j\delta\phi\right]}+O((\delta\phi)^3).
\end{equation}
So, we can see that we can accomplish a fully integration setting
\begin{equation}
    \delta\phi=\delta\phi_0+\int d^Dy\Delta_1(x-y)j(y),
\end{equation}
in the path integral, being
\begin{equation}
\label{eq:d1}
    -\partial^2\Delta_1(x-y)+3\lambda\phi_c^2(x)\Delta_1(x-y)=\delta^D(x-y).
\end{equation}
This equation can be solved exactly writing down
\begin{equation}
   G_n(x)=-\delta^{D-1}(x)\theta(t)\frac{1}{\mu(2^3\lambda)^{\frac{1}{4}}}
   {\rm cn}\left[\left(\frac{\lambda}{2}\right)^{\frac{1}{4}}\mu t+(4n+1)K(i),i\right]
   {\rm dn}\left[\left(\frac{\lambda}{2}\right)^{\frac{1}{4}}\mu t+(4n+1)K(i),i\right]
\end{equation}
when the phase of the exact solution is taken to be $\theta_n=(4n+1)K(i)$ being $K(i)\approx 1.311028777$ an elliptic integral. This identifies an infinite class of quantum field theories for which we are able to compute starting from a subset of infinite countable exact solutions of the classical theory. All such theories are equivalent. So, in the end we have
\begin{equation}
\label{eq:zn}
   Z_n[j]=Z_n[0]e^{i[\int d^Dx j(x)\phi_c^{(n)}(x)+\int d^Dxd^Dyj(x)G_n(x-y)j(y)]}+O((\delta\phi)^3)
\end{equation}
that has the Gaussian form of a free theory. In order to compute the spectrum of the theory, we note that one can write
\begin{equation}
   G_n(x)=-\delta^{D-1}(x)\theta(t)\frac{1}{2\mu^2}\left.
   \frac{d}{du}\phi^{(n)}_c(u,0)\right|_{u=\left(\frac{\lambda}{2}\right)^{\frac{1}{4}}\mu t}. 
\end{equation}
Then
\begin{eqnarray}
\label{eq:prop}
   G_n(x)&=&-\delta^{D-1}(x)\theta(t)\frac{1}{2\mu}
   \left(\frac{2}{\lambda}\right)^\frac{1}{4}\frac{\pi^2}{K(i)^2}\sum_{n=0}^\infty(-1)^n(2n+1)
   \frac{e^{-\left(n+\frac{1}{2}\right)\pi}}{1+e^{-(2n+1)\pi}}\times \nonumber \\
   &&\cos\left((2n+1)\frac{\pi}{2K(i)}\left(\frac{\lambda}{2}\right)^\frac{1}{4}\mu t+\theta_n\right)
\end{eqnarray}
that has the required form
\begin{equation}
   G_n(x)=-\delta^{D-1}(x)\theta(t)\sum_{n=0}^\infty B_n e^{-iM_n t} + c.c.
\end{equation}
with
\begin{equation}
   M_n=(2n+1)\frac{\pi}{2K(i)}\left(\frac{\lambda}{2}\right)^\frac{1}{4}\mu
\end{equation}
that can now be identified with a mass spectrum. This appears like a superimposed spectrum of a harmonic oscillator for a free massive particle. So, at the infrared fixed point the theory takes the form given by eq.(\ref{eq:zn}) that means
\begin{equation}
\label{eq:ms}
   \left.\frac{\delta Z_n}{\delta j(x)}\right|_{j=0,x=0}=\langle\phi(0)\rangle=\mu\left(\frac{2}{\lambda}\right)^\frac{1}{4},
\end{equation}
using ${\rm sn}(K(i),i)=1$, that gives the vacuum expectation value of the field $v$ and we have exploited translational invariance of the vacuum state. This is seen to be different from zero and so behaves exactly as a Higgs field, provided the mass spectrum is the one we gave in eq.(\ref{eq:ms}). The classical solution just represents oscillations around this value.

This analysis can be immediately put at work in the Standard Model given the above formula for the vacuum expectation value. We need to fix a free parameter, $\mu$, and this can be done using the recent discovery of the Higgs boson~\cite{Aad:2012tfa,Chatrchyan:2012ufa}. We fix the mass at $M_H=126\pm 1\ {\rm GeV}$ and assume that is the ground state in the harmonic oscillator spectrum. This gives $\left(\frac{\lambda}{2}\right)^\frac{1}{4}\mu=105\pm 1\ {\rm GeV}$. Similarly, one has $v=246.22\ {\rm GeV}$ with a smaller error. Together we get $\mu=161\pm 1\ {\rm GeV}$ and $\lambda=0.36\pm 0.01$. This gives a mass for the W in agreement with experimental results but, better, we have an estimation for $\lambda$. With respect to the Higgs mechanism, the absence of the mass term changes dramatically the spectrum of the theory while, formally, we get an identical behavior at the tree level. This means that, to distinguish these two approaches, one has to observe excitations of the Higgs field or effects that go at the 1-loop level as, possibly, decay rates into these higher excitations could be strongly depressed as we are going to see.

In order to understand why higher massive states of the Higgs particle are depressed we have to turn our attention to the propagator at the infrared fixed point (\ref{eq:prop}). We note that each state enters with an exponentially smaller weight. These weights will enter into the action of fields of each state into the Gaussian functional. We can formalize this by taking the Fourier transform of the propagator that takes the form after a boost that move our solution to a moving frame
\begin{equation}
    G(p)=\sum_{n=0}^\infty\frac{f_n^2}{p^2-M_n^2+i\epsilon}
\end{equation}
being
\begin{equation}
    f^2_n=(2n+1)^2\frac{\pi^3}{4K^3(i)}\frac{e^{-\left(n+\frac{1}{2}\right)\pi}}{1+e^{-(2n+1)\pi}}
\end{equation}
the weights that we see are exponentially damped. Being this a two-point function, we can interpret the behavior of the theory at the infrared fixed point as an infinite sum of free scalar fields, each one weighing $f_n$. Now, using LSZ reduction, we recognize that the decay rate of one of the Higgs states into a W or Z boson is proportional to $f_n^2$. Then the rates for the decay can be immediately computed as (V=W,Z) \cite{Altarelli:2013tya}
\begin{equation}
   \Gamma_{VV}(n)=\frac{G_FM^3_n}{16\pi\sqrt{2}}\delta_V\beta_V(1-4x+12x^2)
\end{equation}
where $G_F$ the Fermi constant, $\beta_V=\sqrt{1-4x}$ with $x=M^2_V/M^2_n$, $\delta_W=2$, $\delta_Z=1$, $M_n=(2n+1)M_H$ and so $M_H=\frac{\pi}{2K(i)}\left(\frac{\lambda}{2}\right)^\frac{1}{4}\mu$ with all the parameters given above. These would be Standard Model expected rates with increasing masses. The corresponding decay rates one should expect for the impostor field will be
\begin{equation}
   \Gamma^I_{VV}(n)=f_n^2\Gamma_{VV}(n)
\end{equation}
These are the formulas we use in our analysis. 

Then, we give here a table with the first few values of these weights in order to give an idea of the reduction factors for the rates.
\begin{table}[!ht]
\begin{center}
\begin{tabular}{|c|c|c|c|} \hline\hline
n & $f_n^2$        & \%    & \% to SM \\ \hline
0 & 0.6854746582   &  -    &   31     \\ \hline 
1 & 0.2780967321   & 59    &   72     \\ \hline
2 & 0.0333850484   & 95    &   97     \\ \hline
3 & 0.0028276899   & 99.6  &   99.7   \\ \hline
4 & 0.0002019967   & 99.97 &   99.98  \\ \hline
\end{tabular}
\caption{\label{tab:fn} Weights and percentage reductions of the decay rates of Higgs excited states. Percentages are respect to the ground state in the second column and respect to the Standard Model (SM) in the third one.}
\end{center}
\end{table}
It is important to note that the production rate will depend on the mass of the Higgs state we consider, increasing with it. But, due to these large reductions in the rates for the decay of excited states of the Higgs, their observation is increasingly difficult. Firstly, we note that the weight for the decay rate for the fundamental state is not exactly one. Secondly, the number of events for the first excited state, expected at about 315 GeV, is reduced of about sixty percent and so, presently is really difficult to observe. Then, a possible peak seen in the $H\rightarrow ZZ$ at such mass would appear like a statistical fluctuation yet. We also note that bound states of Higgs can form producing also a mass spectrum with even numbers~\cite{Maas:2012tj} so that an excitation at about 250 GeV could be a possibility.

It is interesting to digress on the ground state that has a decay factor of $\mu\approx 0.68$. This means that, with respect to the expectations of the Standard Model, the production rates for the decays $H\rightarrow ZZ$ and $H\rightarrow WW$ are reduced. Indeed, CMS data reach exactly this value but errors are sizable yet. 
Rates are $\mu=0.92\pm 0.28$ for ZZ and $\mu=0.68\pm 0.20$ for WW. These figures are in really close agreement with those we computed in Tab.\ref{tab:fn} but, as said, error bars are too much large to make a claim. Besides, the number of events for ZZ decay seems to be too low yet for this value to settle at a proper value. This must be the same as WW decay in Standard Model and from our expectations. In ATLAS these rates are somewhat higher being $\mu=1.7^{+0.5}_{-0.4}$ and $\mu=1.01\pm 0.31$ respectively, comfortably in agreement with the Standard Model. These figures appear to settle toward the lower value we indicated, for WW decay, as in this case there is a larger number of events and the error bar is large enough. Anyway, excited states, if any, could be observed presently with a lot of difficulties due to the exponential decreasing decay factors and can appear as background yet and so, their production rates are too small with the given luminosity.

From this analysis appears quite clear that a mathematical theory that is consistent exists producing an impostor field for the Higgs boson as originally conceived. Particularly, the only meaningful quantities that could permit to unveil the very nature of the particle recently observed at LHC are the decay rates but, also, eventually observing further heavier states. This latter possibility can be difficult to achieve presently but it could be easier to afford with the upgrade of the LHC where also the nature of the potential of the Higgs field will undergo measurement. 


\end{document}